\documentclass[twocolumn]{aastex7}
\usepackage{soul}
\shorttitle{Near-infrared Variability Detected in Young Star-Forming Dwarf Galaxy SBS~0335-052E}
\shortauthors{Hatano et al.}
\graphicspath{{./}{figures/}}

\begin{document}

\title{Near-infrared Variability Detected in the Young Star-Forming Dwarf Galaxy SBS~0335-052E}

\correspondingauthor{Shun Hatano}
\email{shun.hatano@grad.nao.ac.jp}

\author[0000-0002-5816-4660]{Shun Hatano}
\email{shun.hatano@grad.nao.ac.jp}
\affiliation{National Astronomical Observatory of Japan, 2-21-1 Osawa, Mitaka, Tokyo 181-8588, Japan}
\affiliation{Department of Astronomical Science, The Graduate University for Advanced Studies, SOKENDAI, 2-21-1 Osawa, Mitaka, Tokyo, 181-8588, Japan}

\author[0000-0001-6402-1415]{Mitsuru Kokubo}
\email{mitsuru.kokubo@nao.ac.jp}
\affiliation{National Astronomical Observatory of Japan, 2-21-1 Osawa, Mitaka, Tokyo 181-8588, Japan}
\affiliation{Department of Astronomical Science, The Graduate University for Advanced Studies, SOKENDAI, 2-21-1 Osawa, Mitaka, Tokyo, 181-8588, Japan}

\author[0000-0002-1049-6658]{Masami Ouchi}
\email{masami.ouchi@nao.ac.jp}
\affiliation{National Astronomical Observatory of Japan, 2-21-1 Osawa, Mitaka, Tokyo 181-8588, Japan}
\affiliation{Institute for Cosmic Ray Research, The University of Tokyo, 5-1-5 Kashiwanoha, Kashiwa, Chiba 277-8582, Japan}
\affiliation{Department of Astronomical Science, The Graduate University for Advanced Studies, SOKENDAI, 2-21-1 Osawa, Mitaka, Tokyo, 181-8588, Japan}
\affiliation{Kavli Institute for the Physics and Mathematics of the Universe (WPI), University of Tokyo, Kashiwa, Chiba 277-8583, Japan}

\author[0000-0003-2965-5070]{Kimihiko Nakajima}
\email{knakajima@staff.kanazawa-u.ac.jp}
\affiliation{Institute of Liberal Arts and Science, Kanazawa University, Kakuma-machi, Kanazawa, 920-1192, Ishikawa, Japan}
\affiliation{National Astronomical Observatory of Japan, 2-21-1 Osawa, Mitaka, Tokyo 181-8588, Japan}

\author[0000-0002-3866-9645]{Toshihiro Kawaguchi}
\email{kawaguti@eng.u-toyama.ac.jp}
\affiliation{Graduate School of Science and Engineering, 
University of Toyama, Gofuku 3190, Toyama 930-8555, Japan}

\author[0000-0003-3214-9128]{Satoshi Kikuta}
\email{satoshi.kikuta@nao.ac.jp}
\affiliation{Department of Astronomy, School of Science, The University of Tokyo, 7-3-1 Hongo, Bunkyo, Tokyo 113-0033, Japan}

\author[0000-0001-8537-3153]{Nozomu Tominaga}
\email{nozomu.tominaga@nao.ac.jp}
\affiliation{National Astronomical Observatory of Japan, 2-21-1 Osawa, Mitaka, Tokyo 181-8588, Japan}
\affiliation{Astronomical Science Program, Graduate Institute for Advanced Studies, SOKENDAI, 2-21-1 Osawa, Mitaka, Tokyo 181-8588, Japan}
\affiliation{Department of Physics, Faculty of Science and Engineering, Konan University, 8-9-1 Okamoto, Kobe, Hyogo 658-8501, Japan}

\author[0000-0002-5768-8235]{Yi Xu}
\email{xuyi@icrr.u-tokyo.ac.jp}
\affiliation{Institute for Cosmic Ray Research, The University of Tokyo, 5-1-5 Kashiwanoha, Kashiwa, Chiba 277-8582, Japan}
\affiliation{Department of Astronomy, Graduate School of Science, the University of Tokyo, 7-3-1 Hongo, Bunkyo, Tokyo 113-0033, Japan}

\author[0000-0002-2740-3403]{Kuria Watanabe}
\email{kuria.watanabe@grad.nao.ac.jp}
\affiliation{National Astronomical Observatory of Japan, 2-21-1 Osawa, Mitaka, Tokyo 181-8588, Japan}
\affiliation{Department of Astronomical Science, The Graduate University for Advanced Studies, SOKENDAI, 2-21-1 Osawa, Mitaka, Tokyo, 181-8588, Japan}

\author[0000-0002-6047-430X]{Yuichi Harikane}
\email{hari@icrr.u-tokyo.ac.jp}
\affiliation{Institute for Cosmic Ray Research, The University of Tokyo, 5-1-5 Kashiwanoha, Kashiwa, Chiba 277-8582, Japan}

\author[0000-0001-7730-8634]{Yuki Isobe}
\email{yi264@cam.ac.uk}
\affiliation{Kavli Institute for Cosmology, University of Cambridge, Madingley Road, Cambridge, CB3 0HA, UK}
\affiliation{Cavendish Laboratory, University of Cambridge, 19 JJ Thomson Avenue, Cambridge, CB3 0HE, UK}
\affiliation{Waseda Research Institute for Science and Engineering, Faculty of Science and Engineering, Waseda University, 3-4-1, Okubo, Shin- juku, Tokyo 169-8555, Japan}

\author{Akinori Matsumoto}
\email{matsu@icrr.u-tokyo.ac.jp}
\affiliation{Institute for Cosmic Ray Research, The University of Tokyo, 5-1-5 Kashiwanoha, Kashiwa, Chiba 277-8582, Japan}
\affiliation{Department of Physics, Graduate School of Science, The University of Tokyo, 7-3-1 Hongo, Bunkyo, Tokyo 113-0033, Japan}

\author[0000-0003-4321-0975]{Moka Nishigaki}
\email{moka.nishigaki@grad.nao.ac.jp}
\affiliation{National Astronomical Observatory of Japan, 2-21-1 Osawa, Mitaka, Tokyo 181-8588, Japan}
\affiliation{Department of Astronomical Science, The Graduate University for Advanced Studies, SOKENDAI, 2-21-1 Osawa, Mitaka, Tokyo, 181-8588, Japan}

\author[0000-0001-9011-7605]{Yoshiaki Ono}
\email{ono@icrr.u-tokyo.ac.jp}
\affiliation{Institute for Cosmic Ray Research, The University of Tokyo, 5-1-5 Kashiwanoha, Kashiwa, Chiba 277-8582, Japan}

\author[0000-0003-3228-7264]{Masato Onodera}
\email{monodera@naoj.org}
\affiliation{Subaru Telescope, National Astronomical Observatory of Japan, National Institutes of Natural Sciences (NINS), 650 North A'ohoku Place, Hilo, HI 96720, USA}

\author[0000-0001-6958-7856]{Yuma Sugahara}
\email{sugayu@aoni.waseda.jp}
\affiliation{National Astronomical Observatory of Japan, 2-21-1 Osawa, Mitaka, Tokyo 181-8588, Japan}
\affiliation{Waseda Research Institute for Science and Engineering, Faculty of Science and Engineering, Waseda University, 3-4-1 Okubo, Shinjuku, Tokyo 169-8555, Japan}
\affiliation{Department of Pure and Applied Physics, School of Advanced Science and Engineering, Faculty of Science and Engineering, Waseda University, 3-4-1 Okubo, Shinjuku, Tokyo 169-8555, Japan}

\author[0009-0008-0167-5129]{Hiroya Umeda}
\email{ume@icrr.u-tokyo.ac.jp}
\affiliation{Institute for Cosmic Ray Research, The University of Tokyo, 5-1-5 Kashiwanoha, Kashiwa, Chiba 277-8582, Japan}
\affiliation{Department of Physics, Graduate School of Science, The University of Tokyo, 7-3-1 Hongo, Bunkyo, Tokyo 113-0033, Japan}

\author[0000-0003-3817-8739]{Yechi Zhang}
\email{yczhang@icrr.u-tokyo.ac.jp}
\affiliation{IPAC, California Institute of Technology, 1200 E. California Blvd, Pasadena, CA 91125, USA}

\author[0009-0003-4594-3715]{Ryotaro Chiba}
\email{ryotaro.chiba@grad.nao.ac.jp}
\affiliation{National Astronomical Observatory of Japan, 2-21-1 Osawa, Mitaka, Tokyo 181-8588, Japan}
\affiliation{Department of Astronomical Science, The Graduate University for Advanced Studies, SOKENDAI, 2-21-1 Osawa, Mitaka, Tokyo, 181-8588, Japan}

\author[0000-0003-1169-1954]{Takashi J. Moriya}
\email{takashi.moriya@nao.ac.jp}
\affiliation{National Astronomical Observatory of Japan, 2-21-1 Osawa, Mitaka, Tokyo 181-8588, Japan}
\affiliation{Department of Astronomical Science, The Graduate University for Advanced Studies, SOKENDAI, 2-21-1 Osawa, Mitaka, Tokyo, 181-8588, Japan}
\affiliation{School of Physics and Astronomy, Monash University, Clayton, VIC 3800, Australia}



\begin{abstract}

SBS~0335-052E is a young star-forming dwarf galaxy with a total stellar mass of $M_{*} \lesssim 10^{8}~M_{\odot}$ and an extremely low metallicity ($Z \sim 1/40~Z_{\odot}$), which has long been considered to be devoid of an active galactic nucleus (AGN).
Here we report the detection of temporal flux variability of SBS~0335-052E in near-infrared (NIR) 3-4\ ${\rm \mu}$m bands on timescales of several years, showing dimming and brightening of up to 50\% over 14~years, based on archival data from the Wide-field Infrared Survey Explorer.
Our {spectral energy distribution (SED)} fitting of archival ultraviolet (UV)–NIR photometry, including AGN SED models, indicates that the variable NIR emission arises from an edge-on AGN dust torus. 
The UV–optical emission from the accretion disk is obscured and does not reach us, leading to the dominance of the host galaxy's young stellar population in the UV-optical wavelengths. 
This analysis favors the presence of a Compton-thick, heavily obscured AGN in SBS~0335-052E, consistent with its observed X-ray weakness.
From the SED fitting, we estimate an AGN bolometric luminosity of $L_{\rm bol} = 1.2\times10^{43}\ {\rm erg\ s^{-1}}$, which implies a black hole mass of $M_{\rm BH} \simeq 10^{5}\ M_\odot$ if the AGN is accreting at the Eddington limit.
If confirmed, SBS~0335-052E would be the least massive galaxy known to host an AGN, likely harboring an intermediate-mass black hole.

\end{abstract}



\section{Introduction}

Studies over the past decades have established that a supermassive black hole (SMBH, with a mass of $M_{\rm BH}\sim10^{6-10}\ M_\odot$) resides in almost all massive galaxies 
and the SMBH mass correlates with the host galaxy properties 
\citep[e.g.,][]{1998AJ....115.2285M,2000ApJ...539L..13G,2013ARA&A..51..511K,2013ApJ...764..184M,2015ApJ...813...82R}.
As a possible origin of SMBHs, various pathways of BH formation have been proposed, and they are often summarized into three types of seed BHs: direct collapse of pristine gas clouds, runaway collisions of dense star clusters, and Population~III star remnants \citep[e.g.,][]{1978Obs....98..210R,1984ARA&A..22..471R,2001ApJ...562L..19E,2001ApJ...550..372F, 2020ARA&A..58..257G,2020ARA&A..58...27I}, and these processes are thought to produce seed BHs with masses of order $M_{\rm BH}\sim10^{2-5}~M_\odot$, corresponding to the {intermediate-mass black hole (IMBH) range ($M_{\rm BH}\sim10^{2-5}~M_\odot$)}.
While massive stellar mass BHs with $M_{\rm BH}\lesssim10^2\ M_\odot$ have been detected via gravitational waves \citep[e.g.,][]{2020ApJ...900L..13A,2025arXiv250708219T}, the overall demographics of IMBHs remain unclear.

Extrapolating the well-established empirical relation between host stellar mass $M_{*}$ and SMBH mass $M_{\rm BH}$ to lower stellar masses suggests that IMBHs are hosted by galaxies with $M_{*} \lesssim 10^9\ M_\odot$ \citep[e.g.,][]{2015ApJ...813...82R,2020ARA&A..58..257G}.
Although {\it James Webb Space Telescope (JWST)} observations have recently started witnessing high-redshift ($z \gtrsim 4$) candidates of Active Galactic Nuclei (AGNs) exhibiting broad H$\alpha$ (and sometimes H$\beta$) emission lines in low-mass galaxies with $M_* \sim 10^9~M_\odot$ \citep[e.g.,][]{2023ApJ...954L...4K,2023ApJ...957L...7K,2023A&A...677A.145U,2023arXiv230311946H,2024ApJ...963..129M,2024MNRAS.531..355U,2024A&A...691A.145M,2024Natur.628...57F, 2024ApJ...964...39G, 2024Natur.636..594J, 2025ApJ...986..126K}, IMBH searches {based on the detection of broad H$\alpha$ emission remain challenging at such redshifts due to sensitivity limits in the absence of gravitational lensing} \citep[e.g.,][]{2023ApJ...957L...3P}. 
Therefore, searching for IMBHs in the local universe, which may represent relics of those in the high-redshift universe, is also crucial.
Several attempts have been made to search for low-$z$ IMBH AGN candidates via broad H$\alpha$ line detection, and so far only a handful of them have been confirmed as AGNs with $M_{\rm BH} \sim 10^{5}~M_{\odot}$ hosted in galaxies with $M_{*} \sim 10^{9}~M_{\odot}$ \citep[e.g.,][]{2004ApJ...610..722G,2011ApJ...739...28X,2015ApJ...813...82R,2018ApJ...863....1C,2020ARA&A..58..257G}.

SBS~0335-052E, the target of this study, is a blue compact dwarf and metal-poor galaxy at redshift $z=0.01352$ \citep{1990Natur.343..238I,1992A&A...253...16M}.
SBS~0335-052E has an extremely low stellar mass of $M_* \simeq 10^{7-8}~M_\odot$ \citep{2004A&A...425...51P,2008AJ....136.1415R,2009ApJ...691.1068T}, and is one of the most metal-deficit galaxy known to date \citep[$Z \sim 1/40\ Z_\odot$;][]{1990Natur.343..238I, 1992A&A...253...16M,2001A&A...378L..45I}. 
SBS~0335-052E hosts several compact super star clusters (SSCs), each of which has a stellar mass of $\sim 10^{6}~M_{\odot}$ \citep[e.g.,][]{1997ApJ...477..661T,2008AJ....136.1415R,2009ApJ...691.1068T}.
Its stellar population is extremely young ($\lesssim 10$~Myr; \citealt{2008AJ....136.1415R}), and SBS~0335-052E is undergoing vigorous star formation of $\sim 1~M_{\odot}~\text{yr}^{-1}$ or $\sim 20~M_{\odot}~\text{yr}^{-1}~\text{kpc}^{-2}$ \citep{2008AJ....136.1415R,2009AJ....137.3788J,2017MNRAS.468L..87C} close to the maximum starburst intensity limit of $45~M_{\odot}~\text{yr}^{-1}~\text{kpc}^{-2}$ \citep{1997AJ....114...54M,2009AJ....137.3788J}.
SBS~0335-052E has been extensively studied as a laboratory for young star-forming galaxies since its discovery \citep{1990Natur.343..238I}.

Despite its extremely low stellar mass and metallicity, possible AGN signatures have been reported in SBS~0335-052E.
Spatially-resolved {near-infrared} (NIR) observations with the {\it Hubble Space Telescope (HST)} have revealed that, among the SSCs in SBS~0335-052E, the southern SSCs \citep[referred to as SSC~1 and SSC~2 in the literature;][]{1997ApJ...477..661T} exhibit excess emission at $\gtrsim 2~\mu$m, corresponding to hot thermal dust continua \citep{2004ApJS..154..211H,2008AJ....136.1415R,2008ApJ...678..804E}.
Its  {NIR} colors in the {Wide-field Infrared Survey Explorer ({\it WISE})} $W1$ (3.4~$\mu$m) and $W2$ (4.6~$\mu$m) bands fall in the regime of AGNs ($W1-W2=2.0>0.8$~{in Vega magnitudes}; \citealt{2012ApJ...753...30S}).
A {shallow} 9.7~$\mu$m silicate absorption feature is clearly observed in the {\it Spitzer} mid-infrared spectrum \citep{2004ApJS..154..211H}, as observed in obscured type 2 AGNs \citep[e.g.,][]{2007ApJ...654L..49S,2007ApJ...655L..77H,2015ApJ...803..110H}.
The equivalent widths of its polycyclic aromatic hydrocarbon (PAH) features are unusually low, placing it in the same parameter space as AGNs rather than typical star-forming regions \citep{2002AJ....124.1995P,2004ApJS..154..211H,2005ApJ...633..706W}.
However, such hot dust emission, often seen in low-mass dwarf galaxies, has generally been attributed to peculiar dust distributions associated with intense star formation rather than AGNs \citep[e.g.,][]{2016ApJ...832..119H,2025ApJ...979...36S,2025arXiv251007303R}. 
Observational and theoretical studies suggest that extreme star-forming activity can mimic the hot dust emission typically ascribed to AGNs \citep{2016ApJ...832..119H,2018ApJ...858...38S,2025ApJ...979...36S}.
{In addition, in extremely low-metallicity environments such as SBS~0335-052E, the reduced silicate abundance can produce AGN-like shallow silicate features in star-forming optically thick dusty clouds, while the low carbon abundance limits the formation of PAHs and consequently suppresses their emission \citep[e.g.,][]{2004ApJS..154..211H}.}
Accordingly, hot dust emission of SBS~0335-052E has generally been considered to originate from the intense star formation {\citep{2002AJ....124.1995P,2004ApJS..154..211H,2010ApJ...725.1620A,2014AA...561A..49H,2016ApJ...832..119H,2025arXiv251007303R}.}

Besides the hot dust emission, the optical spectrum of SBS~0335-052E show strong [Ne~{\sc v}] and [Fe~{\sc v}] emission lines, which require high energy photon to photo-ionize the atomic gas \citep[$\gtrsim 50-100$ eV;][]{2001A&A...378L..45I,2009AA...503...61I,2024ApJ...966..170H}. 
To understand the underlying ionizing sources, \cite{2024ApJ...966..170H} estimated ionizing spectrum of SBS~0335-052E from $>$10 optical emission lines, including [Ne~{\sc v}]$\lambda$3426 line,  and 
presented that an ionizing source besides the stellar component is required to explain the observed [Ne~{\sc v}]$\lambda$3426/[Ne {\sc iii}]$\lambda$3869 emission line ratio. 
Interestingly, the X-ray luminosity is small compared to that expected from the estimated ionizing spectrum \citep{2024ApJ...966..170H}.
Because of the high [Ne~{\sc v}]$\lambda$3426/[Ne {\sc iii}]$\lambda$3869 ratio but low X-ray luminosity \citep{2004ApJ...606..213T}, 
X-ray binaries are ruled out as an origin of the [Ne~{\sc v}]$\lambda$3426 emission line \citep{2005ApJS..161..240T,2024ApJ...966..170H}. 
Fast radiative shocks with shock velocity of $\sim 450$ km s$^{-1}$ {were} considered as promising origin \citep{2005ApJS..161..240T}, 
{while} \cite{2012MNRAS.427.1229I} showed that AGNs are not ruled out. 
The estimated ionizing spectral shape and the extreme UV luminosity are consistent with those of the accretion disk of IMBH models.
\cite{2024ApJ...966..170H} also showed that the emission line flux ratio of [Ne~{\sc v}]$\lambda$3426/[Ne~{\sc iii}]$\lambda$3869 is constant within $\sim 1 \sigma -2 \sigma$ for nearly twenty years, suggesting single supernovae event is less likely the origin of the [Ne~{\sc v}]$\lambda$3426 line.\footnote{After the submission of our paper to arXiv, \citet{2025arXiv250207662M} report the detection of the [Ne~{\sc v}]~$\lambda14.32$ emission line in SBS~0335-052E using {\it JWST}/MIRI integral field unit (IFU) observations. They argue that the presence of an IMBH AGN with $M_{\rm BH} \sim 10^{5}~M_{\odot}$ is required to explain the observed high [Ne~{\sc v}]~$\lambda14.32$/[Ne~{\sc ii}]~$\lambda12.8$ ratio, although contributions from very massive stars, ultraluminous X-ray sources, or shocks cannot be completely ruled out as possible ionization sources in the galaxy.}

To robustly test whether SBS~0335-052E hosts an AGN (presumably powered by an IMBH {inferred from its low stellar mass}) as its central engine, in this work we employ {NIR} variability as an independent diagnostic to investigate the origin of the observed hot dust emission.
The stochastic temporal flux variability is a ubiquitous and unique observational property of the AGN emission, observed across a wide range of wavelengths from X-ray to optical and IR on timescales of days to years \citep[for a review, see][]{1997ARA&A..35..445U}.
The X-ray and {ultraviolet (UV)}-optical variability is {thought} to arise from accretion disk instabilities \citep[e.g.,][]{1997ARA&A..35..445U,1998ApJ...504..671K,2009ApJ...698..895K}, whereas the IR variability is due to the reradiation of the dust torus responding to the variable disk UV irradiation \citep[e.g.,][]{1992ApJ...400..502B,2006ApJ...639...46S,2014ApJ...788...48S,2014ApJ...788..159K}.
The stochastic variability is a universal property of AGNs, and thus the detection of variable sources in galactic nuclei can be used to identify AGN candidates.
Indeed, variability-based AGN surveys utilizing time-domain datasets have been observationally confirmed to be a highly effective method for selecting AGNs, particularly low-luminosity AGNs \citep[e.g.,][and references therein]{2011A&A...530A.122P,2018ApJ...868..152B,2020ApJ...896...10B,2020ApJ...894...24K}.
Notably, IR variability enables the identification of dust-obscured AGNs that might evade detection in conventional optical or X-ray observations \citep{2021ApJS..252...32J,2022ApJ...936..104W,2024ApJ...961..211M,2025arXiv250902254H}.

Several studies have searched for AGNs in dwarf galaxies based on optical or IR variability \citep{2020ApJ...896...10B,2022ApJ...936..104W,2020ApJ...900...56S,2023ApJ...945..157H,2024ApJ...975...60A}, but the variability of SBS~0335-052E has not been investigated so far.
In this work, we demonstrate that SBS~0335-052E exhibits clear IR variability and discuss, from several perspectives, the presence of a low-mass AGN, possibly an IMBH, within this galaxy.

Throughout this paper, we adopt the flat $\Lambda$CDM cosmology with parameters of 
$H_{\rm 0} = 70 $ km s$^{-1}$ Mpc$^{-1}$, 
$\Omega_{\rm m} = 0.30$,
and
$\Omega_{\rm \Lambda} = 0.70$.
{Magnitudes are based on the Vega system, and the {\it WISE} magnitudes are converted into fluxes using zero-point fluxes of $f_{\nu,0} = 309.540$ and $171.787$ Jy for the $W1$ and $W2$ bands, respectively\footnote{\href{https://irsa.ipac.caltech.edu/data/WISE/docs/release/All-Sky/expsup/sec4_4h.html}{https://irsa.ipac.caltech.edu/data/WISE/docs/release/All-Sky/expsup/sec4\_4h.html}}
}

\section{Data}
\label{sec:data}

\begin{table}
    \centering
    \begin{tabular}{c c c}
\hline
\hline
\multicolumn{3}{c}{SBS~0335-052E}\\
\hline
MJD [day] & $W1$ [mag] & $W2$ [mag] \\ 
\hline
55232&14.62$\pm$ 0.03& 12.56$\pm$ 0.02\\ 
55422&14.54$\pm$ 0.03& 12.51$\pm$ 0.01\\
56695&14.55$\pm$ 0.02& 12.62$\pm$ 0.02\\
56885&14.53$\pm$ 0.03& 12.61$\pm$ 0.02\\
57056&14.52$\pm$ 0.02& 12.63$\pm$ 0.02\\
57249&14.49$\pm$ 0.02& 12.60$\pm$ 0.02\\
57416&14.57$\pm$ 0.03& 12.57$\pm$ 0.02\\
57613&14.51$\pm$ 0.03& 12.58$\pm$ 0.02\\
57779&14.53$\pm$ 0.03& 12.56$\pm$ 0.02\\
57978&14.48$\pm$ 0.02& 12.46$\pm$ 0.02\\
58138&14.46$\pm$ 0.02& 12.41$\pm$ 0.02\\
58344&14.35$\pm$ 0.03& 12.24$\pm$ 0.02\\
58503&14.37$\pm$ 0.02& 12.24$\pm$ 0.01\\
58708&14.37$\pm$ 0.03& 12.19$\pm$ 0.02\\
58867&14.33$\pm$ 0.02& 12.18$\pm$ 0.02\\
59075&14.32$\pm$ 0.03& 12.22$\pm$ 0.02\\
59232&14.37$\pm$ 0.02& 12.21$\pm$ 0.01\\
59440&14.36$\pm$ 0.03& 12.22$\pm$ 0.02\\
59599&14.43$\pm$ 0.03& 12.29$\pm$ 0.02\\
59804&14.43$\pm$ 0.02& 12.25$\pm$ 0.02\\
59963&14.40$\pm$ 0.02& 12.29$\pm$ 0.02\\
60171&14.41$\pm$ 0.03& 12.36$\pm$ 0.02\\
60329&14.48$\pm$ 0.02& 12.29$\pm$ 0.02\\
\hline
\hline
    \end{tabular}
        \caption{{\it WISE} photometry of SBS~0335-052E corrected for the systematic offsets.  
        }
    \label{tab:photometry_WISE}
\end{table}

We use multi-epoch $W1$ and $W2$ band NIR photometric data taken by the {\it Wide-field Infrared Survey Explorer} \citep[\textit{WISE},][]{2010AJ....140.1868W} to explore the time variability in SBS~0335-052E.
The W1 and W2 bands correspond to NIR wavelengths of $3.4~\mu$m and $4.6~\mu$m, respectively.
{\it WISE} acquired all-sky NIR images, including the $W1$ and $W2$ band data, {around} 2010 by the mission dubbed AllWISE \citep{2010AJ....140.1868W}, and extended the mission dubbed NEOWISE in the post-cryo phase for the $W1$ and $W2$ band data twice a year on average since 2013 until 2024 \citep{2011ApJ...731...53M,2014ApJ...792...30M}.
The time baseline of the $W1$ and $W2$ band data released spans 14 years (NEOWISE Final Data Release).

%
%
We retrieve multi-epoch $W1$ and $W2$ band photometry of SBS~0335-052E from the WISE All-Sky and NEOWISE Single Exposure (L1b) Source Tables available at the Infrared Processing and Analysis Center (IPAC) Infrared Science Archive (IRSA) using a cone search radius of 2 arcsec.
The photometric data of SBS~0335-052E were obtained in 23 visits, 
2 of which were covered by AllWISE in 2010 and the remaining 21 by NEOWISE in 2013-2024.
Each visit consists of $\gtrsim 10$ frames taken in several days.
We calculate a weighted mean and standard error of the multiple magnitude measurements in each visit, removing erroneous measurements with standard photometry flags\footnote{\href{https://wise2.ipac.caltech.edu/docs/release/neowise/expsup/sec2_3.html}{https://wise2.ipac.caltech.edu/docs/release/neowise/expsup/sec2\_3.html}}.
We obtain a total of 23 photometric measurements and errors for each band in the epoch of 2010-2024.

The $W1$ and $W2$ photometry may include systematic offsets 
due to time-dependent and position-dependent flux zero-point determination \citep{2023AJ....165...36M}.
We evaluate the systematic offsets of the {\it WISE} $W1$- and $W2$-band multi-epoch photometry relative to the AllWISE Source Catalog photometry. 
In the catalog, SBS~0335–052E is recorded with magnitudes of $W1 = 14.517$~mag and $W2 = 12.528$~mag.
Using these values as references, we searched for AllWISE sources within 3600 arcsec of SBS~0335–052E whose magnitudes fall within ±0.5 mag of these values. 
We found 1312 such AllWISE sources in the $W1$ band and 270 in the $W2$ band.
We cross-match these AllWISE sources with the NEOWISE single-epoch photometry table (L1b), and create a light curve for each object.
We average the light curves normalized with $W1$ and $W2$ magnitudes taken from the AllWISE Source Catalog, and obtain the systematic offsets. 
We find that the systematic offsets are small, within 0.03~mag and 0.05~mag for the $W1$ and $W2$ bands, respectively.
{\it WISE} photometry of SBS~0335-052E corrected for the systematic offsets is tabulated in Table~\ref{tab:photometry_WISE}.

\section{Results}
\label{sec:result}

\begin{figure*}[t]
\plotone{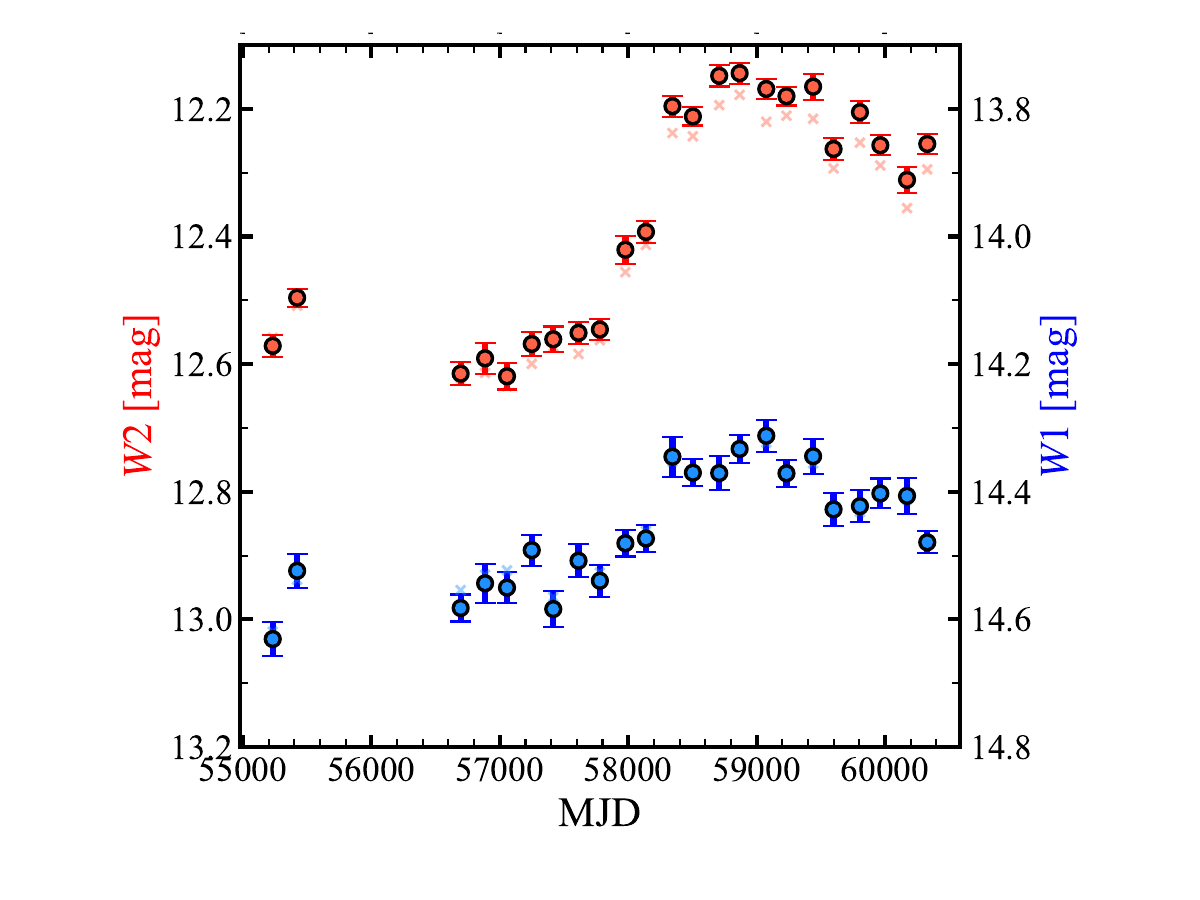}
\caption{{\it WISE} W1 and W2 band light curves of SBS~0335-052E. Blue and red circles indicate the $W1$ and $W2$ band magnitudes, respectively. Crosses denote the light curve data before correction for systematic offsets.}
 \label{fig:time_var}
\end{figure*}


Figure~\ref{fig:time_var} shows the {\it WISE} $W1$ and $W2$ band light curves of SBS~0335-052E corrected for the systematic offsets (Table~\ref{tab:photometry_WISE}).
{
Figure~\ref{fig:time_var} reveals, for the first time, clear NIR variability in SBS~0335-052E.
The $W2$ band brightened by about 0.4~mag between Modified Julian Date (MJD) $\sim$ 57000 and 59000, corresponding to a luminosity increase of roughly 50\%.
A similar trend is seen in the $W1$ band, which brightened by about 0.2~mag (approximately a 20\% increase in luminosity) over the same period.
These variations are highly significant, with detection levels of $\gtrsim 20\sigma$ in $W2$ and $\gtrsim 6\sigma$ in $W1$.
}
{
Prior to this large brightening, the $W2$ band experienced an earlier bright phase followed by a significant dimming from MJD $\sim$55600 to 56800, detected at the $\sim 7\sigma$ level.
In addition to these multi-year variations, we also detect higher-frequency variability on roughly 1-year timescales, corresponding to fluctuations at the 2–3$\sigma$ level around MJD $\sim$58900 in the $W2$ band.
}

The standard deviations of the W1 and W2 band fluxes are 0.037~mJy and 0.283~mJy, respectively, and the single-temperature black body fitting to this variable component reveals the black body temperature of $\sim 380$~K and the black body luminosity of $L_{\rm BB} \sim 4\times 10^{41}~{\rm erg~s}^{-1}$ (see Section~\ref{sec:appendix} for more details about the dust temperature).
Such bright, long-duration luminosity variations with multiple rising and declining phases cannot be attributed to transient one-off events. 
In contrast, AGN variability is stochastic, and such variations are commonly observed in the NIR bands \citep{1989ApJ...337..236C,2004MNRAS.350.1049G,2004ApJ...600L..35M,2019ApJ...886...33L,2020ApJ...900...58Y,2022MNRAS.516.2876M,2022Univ....8..304L}. 
Therefore, the luminosity variations seen in Figure~\ref{fig:time_var} strongly suggest that SBS~0335-052E (presumably its IR-brightest SSC~1+2 region) hosts an AGN.
Other possible explanations will be discussed in Section~\ref{sec:other_scenario}.

\section{Discussion}

\subsection{Origin of the NIR variability}

\subsubsection{SED modelling {under} the AGN scenario}
\label{sec:AGN_scenario}

\begin{figure*}[t]
\plotone{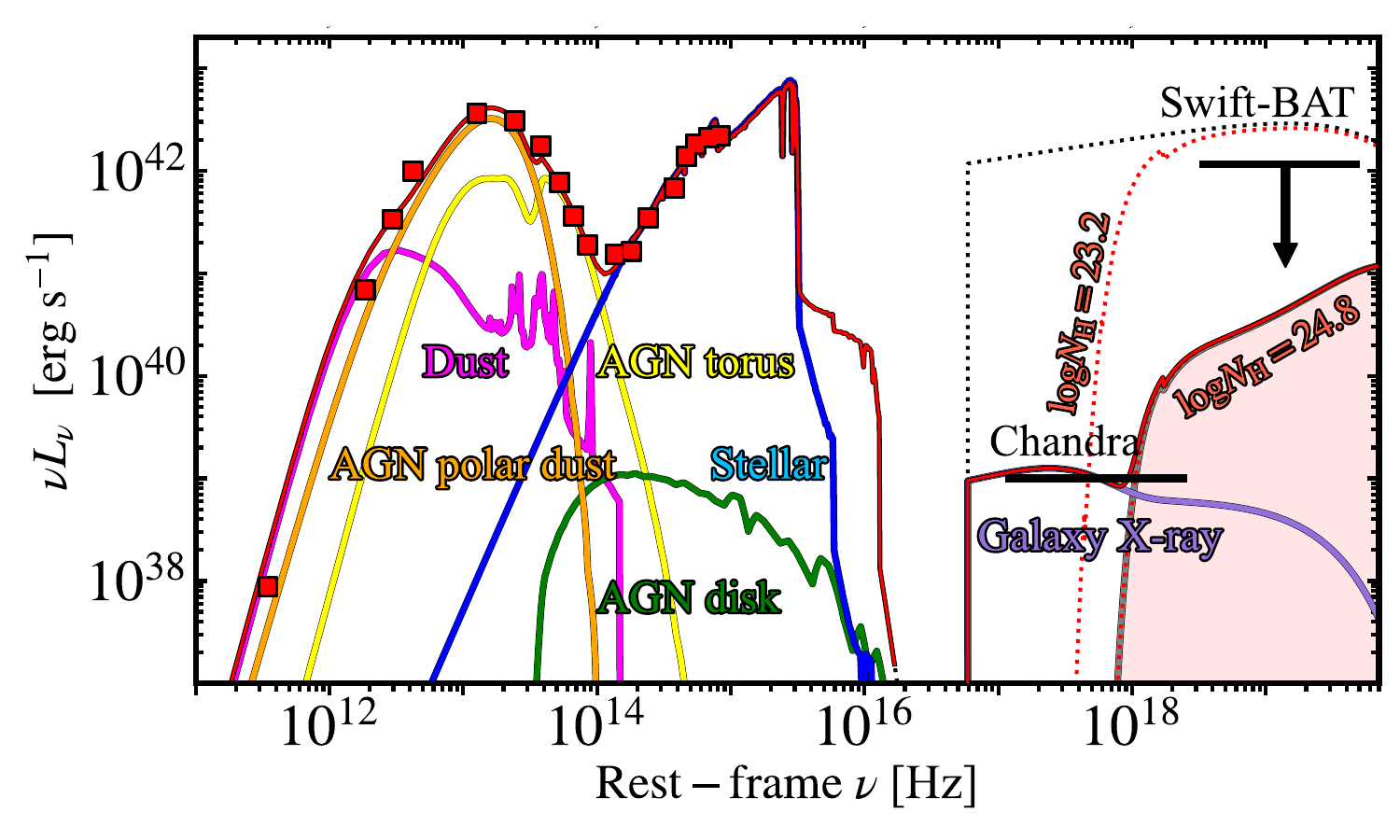}
\caption{X-ray-to-radio SED of SBS~0335-052E. 
The red squares and the black horizontal lines indicate the photometry of SBS~0335-052E (Table~\ref{tab:photometry}). The red line at $\nu \lesssim 10^{16}$ Hz represents the best-fit {\tt CIGALE} SED model. The contributions from dust in the host galaxy, the AGN dust torus, AGN polar dust, the AGN accretion disk, and the stellar population are shown by the magenta, yellow, orange, green, and blue curves, respectively (Section~\ref{sec:AGN_scenario}). The dotted black line indicates the intrinsic X-ray spectrum predicted from the best-fit SED (Section~\ref{sec:appendix}).
The red dotted and grey solid lines present the net transmitted X-ray spectrum calculated from the intrinsic X-ray spectrum with the hydrogen column density of $\log (N_{\rm H}/{\rm cm^{-2}})=23.2$ (with $Z=Z_{\odot}$) and $24.8$ (with $Z=Z_{\odot}/40$), respectively. 
The red shade indicates the net transmitted AGN X-ray spectrum range allowed if we take $\log (N_{\rm H}/{\rm cm^{-2}})=24.8$ (with $Z=Z_{\odot}/40$) as the lower limit of $N_{\rm H}$.
The purple curve indicates the model of the unobscured galaxy's X-ray emission, and the red solid curve at $\gtrsim 10^{16}$ Hz represents the sum of the unobscured galaxy X-ray emission and the transmitted X-ray emission for the hydrogen column density of $\log (N_{\rm H}/{\rm cm^{-2}})=24.8$ (with $Z=Z_{\odot}/40$).}
 \label{fig:cigale}
\end{figure*}

\begin{table}
\centering
\begin{tabular}{c c c}
\hline
\hline
\multicolumn{3}{c}{SBS~0335-052E}\\
\hline
wavelength [$\mu$m] & flux [mJy] & Reference \\ 
\hline
0.360 & 0.73 $\pm$ 0.021 & 1, 5\\
0.440 & 0.73 $\pm$ 0.013 & 1, 5\\
0.550 & 0.85 $\pm$ 0.015 & 1, 5\\
0.641 & 0.81 $\pm$ 0.007 & 1, 5\\
0.791 & 0.49 $\pm$ 0.004 & 1, 5\\
1.25  & 0.40 $\pm$ 0.023 & 2, 5\\
1.65  & 0.25 $\pm$ 0.0493& 2, 5\\
2.20  & 0.31 $\pm$ 0.1320& 2, 5\\
3.55  & 0.62 $\pm$ 0.3890& 3, 5\\
4.49  & 1.50 $\pm$ 0.0493& 3, 5\\
5.73  & 4.07 $\pm$ 0.1320& 3, 5\\
7.87  &12.73 $\pm$ 0.3890& 3, 5\\
12.3  &   35 $\pm$ 6     & 4, 5\\
23.7  & 79.0 $\pm$ 3.12  & 3, 5\\
71.1  & 64.4 $\pm$ 5.7   & 5\\
101.2 & 31.3 $\pm$ 4.75  & 5\\
162.7 & 10.4 $\pm$ 3.5   & 5\\
866.5 & 0.07 $\pm$ 0.070 & 5\\
\hline
\hline
band [keV] & luminosity [erg s$^{-1}$] & Reference\\
\hline
0.5--10.0 & $3.5\times10^{39}$ & 6 \\
14--195 & $< 3.0\times 10^{42}$& This work \\
\hline
    \end{tabular}
        \caption{Multi-wavelength photometry of SBS~0335-052E. In the reference column, (1)–(6) correspond to \cite{1998A&A...338...43P}, \cite{2000A&A...363..493V}, \cite{2008ApJ...678..804E}, \cite{2001AJ....122.1736D}, \cite{2014AA...561A..49H}, and \cite{2004ApJ...606..213T}, respectively. Photometry from 1 to 10~$\mu$m is corrected for nebular continuum emission, as described in \cite{2014AA...561A..49H}. Optical fluxes have been corrected for Galactic extinction.
        }
    \label{tab:photometry}
\end{table}

\begin{deluxetable*}{lc}
\label{tab:cigale_parameters}
\tablewidth{0pt}
\tablecaption{{\tt CIGALE} parameters.}
\tablehead{Parameter & Value\\
}
\startdata
\hline
\hline
\multicolumn{2}{c}{Star formation history}\\
\multicolumn{2}{c}{({\tt sfhdelayed} module)} \\
\hline 
$e$-folding time of the main stellar population, {\tt{tau\_main}} (Myr) & 1, {\bf 10},  100 \\
age of the main stellar population, {\tt{age\_main}} (Myr) & {\bf 100} \\
\hline
\multicolumn{2}{c}{Single stellar population}\\
\multicolumn{2}{c}{({\tt bc03} module; \citealt{2003MNRAS.344.1000B})}\\
\hline
Metallicity & {\bf 0.0001}, 0.0004 \\
\hline
\multicolumn{2}{c}{Dust attenuation}\\
\multicolumn{2}{c}{({\tt dustatt\_powerlaw} module; \citealt{2019AA...622A.103B})}\\
\hline
Av\_young & 0.01, 0.03, {\bf 0.05}, 0.1 \\
uv\_bump\_amplitude & 0.0, {\bf 0.75} \\
\hline
\multicolumn{2}{c}{Galactic dust emission}\\
\multicolumn{2}{c}{({\tt dale2014} module; \citealt{2014ApJ...784...83D})}\\
\hline
\hline
\multicolumn{2}{c}{AGN emission}\\
\multicolumn{2}{c}{({\tt skirtor2016} module; \citealt{2012MNRAS.420.2756S,2016MNRAS.458.2288S})}\\
\hline
Average edge-on optical depth at 9.7 micron & 7,9,{\bf 11} \\
Index that sets dust density gradient with polar angle & {\bf 1.5} \\
Ratio of outer to inner torus radius & {\bf 10} \\
Inclination angle $i$& 0, 10, 20, 30, 40, 50, 60, 70, 80, {\bf 90} \\
AGN fraction in total IR luminosity & 0.1, 0.3, 0.7, 0.8, 0.9, {\bf 0.95} \\
$E(B-V)$ of polar dust & 0.05, {\bf 0.1} \\
Polar dust temperature & 100, {\bf 200}, 300 \\
\hline
\enddata
\tablecomments{The best-fit values of the parameters are shown in bold. Parameter ranges that are not specified in this table are set to the default values of {\tt CIGALE}.}
\end{deluxetable*}

The variable hot dust emission observed in SBS~0335-052E cannot be naturally explained by standard galaxy stellar and dust emission models and {hence} strongly suggests the presence of an AGN. 
Here, we perform spectral energy distribution (SED) modelling to demonstrate that the overall SED of SBS~0335-052E, including the hot dust emission, can be well accounted for within the AGN scenario.

We plot the broad-band SED of SBS~0335-052E in Figure~\ref{fig:cigale}, where the photometric measurements of SBS~0335-052E in the optical to sub-millimeter bands are taken from \cite{2014AA...561A..49H} (see Table~\ref{tab:photometry} for details).
The photometry from 1 to 10~$\mu$m is corrected for nebular continuum emission, as described in \cite{2014AA...561A..49H}. 
Note that the UV-optical-IR photometric data include contributions from both the SSCs and the diffuse stellar component \citep[e.g.,][]{2004A&A...425...51P,2008AJ....136.1415R}, so the following SED analysis reflects the galaxy-wide, averaged stellar populations.

Figure~\ref{fig:cigale} and Table~\ref{tab:photometry} also include X-ray measurements.
In the hard X-ray band, we place an upper limit of $< 8.40 \times 10^{-12}$~erg~s${}^{-1}$~cm${}^{-2}$ ($< 3.0\times 10^{42}~{\rm erg~s}^{-1}$) in the 14–195 keV band, based on the non-detection of SBS~0335-052E in the 105-Month {\it Swift}-Burst Alert Telescope (BAT) All-sky Hard X-Ray Survey \citep{2018ApJS..235....4O}.
In the soft X-ray band, the {\it Chandra} data show X-ray emission of $(3.5-4.3)\times 10^{39}$ erg s$^{-1}$ at 0.5--10.0 keV \citep{2004ApJ...606..213T}.

In this study, we fit the SED of SBS~0335–052E using {\tt CIGALE} v2022.1 \citep{2019A&A...622A.103B,2020MNRAS.491..740Y,2022ApJ...927..192Y}, including AGN models, and determine the best-fit model.
Since {\tt CIGALE} requires absorption-corrected X-ray fluxes \citep{2020MNRAS.491..740Y,2022ApJ...927..192Y}, {which are not directly available from observations for this object}, the fitting is performed over the optical to sub-millimeter bands without using the X-ray module (see Section~\ref{sec:appendix} for a discussion of the X-ray emission).
We summarize the modules and parameter ranges used in the {\tt CIGALE} SED fitting in Table~\ref{tab:cigale_parameters}, and present the best-fit model in Figure~\ref{fig:cigale}.

The best-fit SED, which is composed of a young stellar population and a heavily-obscured AGN, reproduces the overall shape of the SBS~0335-052E SED, being dominated by the cold dust at $\sim 10^{12-13}$ Hz, the AGN polar dust at $\sim 10^{13-14}$ Hz, and stars at $\sim 10^{14-15.5}$ Hz with a negligible contribution of the obscured AGN accretion disk.
The NIR to mid-infrared emission of SBS~0335-052E is naturally explained by the dust torus emission of the obscured AGN.
Note that the heavy obscuration toward the AGN in SBS~0335-052E is consistent with the 9.7~$\mu$m silicate absorption feature observed in the {\it Spitzer} mid-infrared spectrum reported by \cite{2004ApJS..154..211H}.

With this best-fit {\tt CIGALE} SED model, we obtain the intrinsic AGN bolometric luminosity as $L_{\rm bol} = 1.21\times 10^{43}$ erg s$^{-1}$, and the best-fit total stellar mass as $M_{*} = 9.14~(\pm 0.05)\times 10^7 \ M_\odot$.

As shown above, the NIR emission in SBS~0335-052E is naturally explained by the heavily-obscured AGN model. 
Thus, as is commonly observed in nearby type~2 (obscured) AGNs \citep[e.g.,][]{2004MNRAS.350.1049G,2020MNRAS.495.2921N,2022MNRAS.516.2876M}, the variability of the dust torus NIR emission in SBS~0335-052E can be interpreted as a response to the stochastically variable UV-optical emission from the accretion disk, with the disk emission itself {hidden} from direct view.

\subsubsection{Line-of-Sight Dust Extinction and Suppression of AGN X-ray Emission}
\label{sec:appendix}

\begin{figure}[t]
\plotone{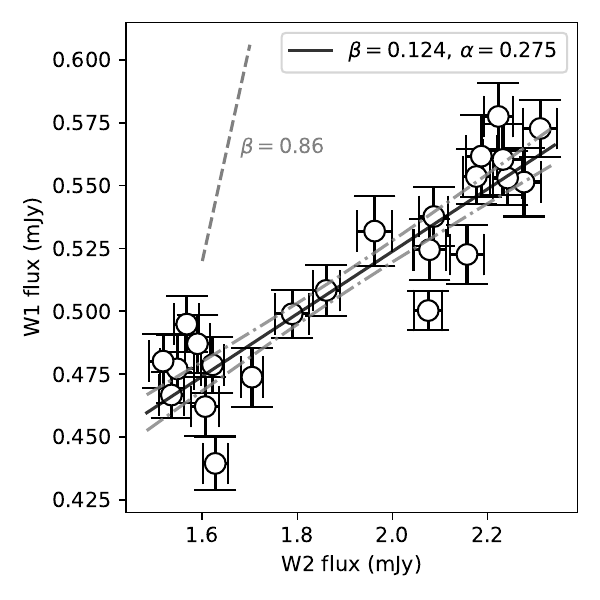}
\caption{{\it WISE} W1 and W2 band fluxes of SBS~0335-052E (Circles). The solid line is the best-fit linear regression line ($y=\alpha + \beta \times x$), and the dot-dashed lines indicate the $16-84$\% percentile range of the regression. {The dashed line shows a reference slope of $\beta = 0.86$, which is a typical flux–flux slope for unobscured AGNs \citep{2022MNRAS.516.2876M}.}}
\label{fig:fluxflux}
\end{figure}

In Section~\ref{sec:AGN_scenario}, we showed through SED model fitting that the putative AGN in SBS~0335-052E is heavily obscured. Here, we estimate the line-of-sight dust extinction of the variable hot dust component using an independent approach: the two-band flux-variation gradient (FVG) method \citep[e.g.,][]{1981AcA....31..293C,1992MNRAS.257..659W,2004MNRAS.350.1049G}.
Following \cite{2022MNRAS.516.2876M}, who first applied the FVG method to {\it WISE} W1- and W2-band light curves of type 1 and type 2 AGNs, we use their approach to derive the visual extinction $A_{V}$ toward the AGN in SBS~0335-052E.
We then compare the inferred dust extinction and gas column density with the {\it Chandra} and {\it Swift} X-ray measurements \citep{{2004ApJ...606..213T}}, showing that the SED modelling in the AGN scenario provided in Section~\ref{sec:AGN_scenario} is consistent with the observed weak X-ray emission in SBS~0335-052E.

The idea of the IR FVG is that the variable hot dust emission in AGNs should have a uniform color temperature of $\sim$1,000~K, set by the dust sublimation temperature \citep{1992ApJ...400..502B,2014ApJ...784L..11Y,2022MNRAS.516.2876M}. 
\cite{2022MNRAS.516.2876M} empirically determined the $W1-W2$ color of the variable component in unobscured type~1 AGNs as $\beta \equiv \Delta W1/\Delta W2 = 0.86 \pm 0.10$ (corresponding to $\sim$1,000~K), based on flux variation gradients measured in flux–flux plots of the $W1$ and $W2$ band {\it WISE} light curves.
\cite{2022MNRAS.516.2876M} observe that obscured AGNs exhibit smaller $\beta$ than 0.86, which indicates the presence of chromatic dust extinction affecting the NIR hot dust emission.

Figure~\ref{fig:fluxflux} shows {\it WISE} $W1$ and $W2$ fluxes for all of the observing epochs. 
Following \cite{2022MNRAS.516.2876M}, the $W1-W2$ color of the variable component in SBS~0335-052E, represented by flux variation gradient $\Delta W1/\Delta W2$, is evaluated by fitting a linear regression line ($y=\alpha + \beta \times x$) to the flux-flux plot of the $W1$ and $W2$ band light curves (Figure~\ref{fig:fluxflux}).
From the linear regression analysis with Josh Meyers' Python port of {linmix\_err} \citep{2007ApJ...665.1489K}, we obtain (Figure~\ref{fig:fluxflux}):
\begin{eqnarray}
\beta = \frac{\Delta W1}{\Delta W2} = 0.124~(\pm 0.015).
\end{eqnarray}

By comparing with the typical value for unobscured AGNs of $\beta = 0.86 \pm 0.10$ \citep{2022MNRAS.516.2876M}, we get a color excess between the $W1$ and $W2$ bands of SBS~0335-052E as $E(W1-W2) = A_{W1} - A_{W2} = 2.103 \pm 0.130~\text{mag}$.
This color excess corresponds to a visual extinction of $A_{V} = 109 \pm 9$ mag, assuming Fitzpatrick’s dust extinction curve \citep{1999PASP..111...63F}.
The high visual extinction indicates that AGN in SBS~0335-052E is viewed nearly edge-on and thus heavily obscured, which is consistent with the result of the {\tt CIGALE} SED fitting (Table~\ref{tab:cigale_parameters}).

Assuming a metallicity-dependent gas-to-dust ratio of $N_{\rm H}/A_{V} \simeq 10^{21.2}(Z/{Z_{\odot}})^{-1} \text{cm}^{-2} \text{mag}^{-1}$ \citep[e.g.,][]{1995A&A...293..889P,2012ApJ...759...95N,2022MNRAS.516.2876M}, we obtain hydrogen column densities as $\log (N_{\rm H}/{\rm cm}^{-2}) \sim 23.2$ and $24.8$ for $Z=Z_{\odot}$ and $Z=Z_\odot/40$, respectively. 
We consider these two metallicities to account for the possibility that the gas in the vicinity of the AGN may be more metal-enriched than the galaxy-wide average, and thus its metallicity remains uncertain.
We note that the estimated hydrogen column densities are lower limits of the actual hydrogen gas column densities because a large fraction of the $N_{\rm H}$ measured in X-ray-obscured AGNs is {thought} to be contributed from dust-free gas particles inside the dust sublimation radius of the dust torus, namely the broad line region (BLR) \citep{2022MNRAS.516.2876M}.
The large hydrogen column density implies that SBS~0335-052E is a heavily obscured AGN, being consistent with the SED analysis in Section~\ref{sec:AGN_scenario}.

To estimate the AGN X-ray flux expected based on the gas column density derived above, we employ the {\tt CIGALE} best-fit SED model supplemented with the {\tt CIGALE} X-ray module \citep{2022ApJ...927..192Y} and apply the corresponding X-ray extinction to the model.
The intrinsic AGN X-ray emission model is presented in Figure~\ref{fig:cigale}, in which the black dotted curve indicates the intrinsic X-ray emission predicted from the best-fit model with the AGN photon index, exponential cutoff energy, optical-to-X-ray spectral index, and AGN X-ray angle coefficients are fixed to $\Gamma=1.8$, $E_{\rm cut}=300~{\rm keV}$, $\alpha_{\rm ox} = -1.1$, and $(a_{1}, a_{2}) = (0.5, 0)$, respectively.
$\alpha_{\rm ox}$ is fixed to the value predicted by the $\alpha_{\rm ox}$–$L_{2500,\text{\AA}}$ relation of \cite{2007ApJ...665.1004J}, using the AGN intrinsic monochromatic luminosity at 2500\AA~($L_{2500,\text{\AA}}$) derived from the best-fit SED of SBS~0335-052E.
As evident from Figure~\ref{fig:cigale}, in the absence of any X-ray attenuation, the AGN X-ray emission model would dramatically overpredict the X-ray luminosity, in complete contradiction to the {\it Chandra} and {the} {\it Swift}-BAT X-ray non-detections.

Then, to convert the intrinsic AGN X-ray emission predicted by the {\tt CIGALE} model into the {attenuated}  X-ray emission, we apply attenuation due to both photoelectric absorption and Thomson scattering.
To calculate the total opacities for the two cases of $N_{\rm H}$ mentioned above, namely $\log(N_{\rm H}/{\rm cm}^{-2}) = 23.2$ (for $Z = Z_{\odot}$) and $\log(N_{\rm H}/{\rm cm}^{-2}) = 24.8$ (for $Z = Z_{\odot}/40$), we use {\tt zvphabs} (photoelectric absorption) and {\tt cabs} (Thomson scattering) models in {\tt XSPEC} v12.14.1.
In {\tt zvphabs}, the $Z$-dependent photoelectric absorption cross section per hydrogen atom, $\sigma_{\rm ph}(Z)$, as given in \cite{1996ApJ...465..487V}, is adopted.
Because photoelectric absorption in the X-ray band arises primarily from metals, the optical depth $N_{\rm H}\sigma_{\rm ph}(Z)$ depends on $Z$, which can be incorporated directly in {\tt zvphabs}.
The optical depth due to Thomson scattering is proportional to $N_{\rm H}\sigma_{T}$.
The attenuated X-ray emission models are then calculated using the total opacities for $\log(N_{\rm H}/{\rm cm}^{-2}) = 23.2$ (for $Z = Z_{\odot}$) and $\log(N_{\rm H}/{\rm cm}^{-2}) = 24.8$ (for $Z = Z_{\odot}/40$), as shown in Figure~\ref{fig:cigale}.


In Figure~\ref{fig:cigale}, the AGN model with the Compton-thin obscuration of $\log (N_{\rm H}/{\rm cm}^{-2}) = 23.2$ clearly overpredicts the soft X-ray emission and is inconsistent with the observed {\it Chandra} X-ray luminosity.
If we instead adopt the Compton-thick obscuration with $\log (N_{\rm H}/{\rm cm}^{-2}) = 24.8$, essentially no soft X-ray AGN emission is expected to be observed.
As noted above, the actual gas column density could be even higher than $\log (N_{\rm H}/{\rm cm}^{-2}) = 24.8$ due to the contribution from the dust-free BLR gas; in that case, any soft X-ray photons with $E \lesssim 10$~keV would be completely absorbed before reaching us.

In this Compton-thick heavily obscured AGN scenario, we include an additional unobscured X-ray emission component from low-mass X-ray binary (LMXB) and high-mass X-ray binary (HMXB) populations to reproduce the observed {\it Chandra} X-ray luminosity.
This component is also implemented using the X-ray module of {\tt CIGALE}.
The LMXB and HMXB parameters are fixed to $\delta_{\rm LMXB}=-1.3$ and $\delta_{\rm HMXB}=-1.3$, where $\delta_{\rm LMXB}$ and $\delta_{\rm HMXB}$ represent the deviations from the empirical scaling relations of LMXB X-ray luminosity versus stellar mass ($M_{*}$) and HMXB X-ray luminosity versus star-formation rate (SFR), respectively \citep{2019ApJS..243....3L,2021ApJ...907...17L}.
These parameters, defined in Equation~4 of \cite{2022ApJ...927..192Y}, are chosen to match the observed X-ray luminosity of SBS~0335-052E \citep[note that deviations of $\gtrsim$1~dex are common in very low-SFR and/or low-$M_{*}$ galaxies;][]{2022ApJ...927..192Y}.

The combined X-ray model, consisting of a Compton-thick ($\log (N_{\rm H}/{\rm cm}^{-2}) = 24.8$) obscured AGN and an unobscured X-ray binary populations, is shown as the red curve in Figure~\ref{fig:cigale}.
This model provides our proposed description of the X-ray spectrum of SBS~0335-052E and is consistent with both the {\it Chandra} detection of the weak soft X-ray emission and the {\it Swift}-BAT hard X-ray upper limit.
It is also compatible with evidence that the {\it Chandra}-detected soft X-ray emission is spatially extended \citep{2004ApJ...606..213T,2018MNRAS.480.1081K}, as well as with the possibility that the observed soft X-ray component at $E<8$~keV can be explained by an unobscured power-law spectrum \citep[$\log (N_{\rm H}/{\rm cm}^{-2}) \sim 21.8$;][]{2004ApJ...606..213T}.

\subsubsection{Other possible scenario}
\label{sec:other_scenario}

A subset of core-collapse supernovae (SNe) emit IR thermal radiation due to either IR echo emission from dust grains in the surrounding circumstellar medium (CSM) or IR emission resulting from dust formation within the SN ejecta \citep[e.g.,][]{1983ApJ...274..175D,2016ApJ...833..231T,2019ApJS..241...38S,2021ApJ...919...17S,2024ApJ...977...98V}. 
In particular, Type IIn SNe, which exhibit strong ejecta-CSM interaction \citep[see][for a review]{2017hsn..book..403S}, can show exceptionally luminous IR dust thermal emission light curves characterized by a slow evolution over several years \citep[e.g.,][]{2003AJ....126.1939S,2004MNRAS.352..457P,2011ApJ...741....7F,2014Natur.511..326G,2014ApJ...797..118F,2016ApJ...816L..13F,2019ApJ...872..135K}. 
The luminous IR dust thermal emission has also been observed in optically elusive Type IIn SNe \citep[e.g.,][]{2017ApJ...837..167J,2018ApJ...863...20J}.

The IR emission observed in SBS~0335–052E could, in terms of IR luminosity, potentially be explained by SN-related dust emission; however, its color temperature and temporal behavior reveal clear differences from typical SN-related dust IR emission.
Newly formed dust in SN ejecta show high temperature near $\sim 1000 $ K (e.g., \citealt{2004MNRAS.352..457P,2019ApJ...872..135K}), which is higher compared to those of SBS~0335-052E $\sim$800~K \citep{2008AJ....136.1415R}.
Also, whereas a smooth rise and decline would be expected in the IR light curve of a SN, SBS~0335-052E actually exhibits a dimming between MJD $=$ 55600 and 56800 prior to the major brightening event (Figure \ref{fig:time_var}). 
We note that SN precursors \citep[e.g.,][]{2013MNRAS.430.1801M,2014ApJ...780...21M,2014ApJ...789..104O,2021ApJ...907...99S}, phenomena that occur ahead of a terminal SN explosion, may explain the dimming observed between MJD $=$ 55600 and 56800.
Although we do not completely rule out the possibility that the NIR variability of SBS~0335-052E originates from the combination of SN precursor events and terminal SN~IIn explosion, explaining both the dimming and long-lasting {NIR} variability is difficult overall.

Recently, \citet{2025arXiv250803912P} proposed that the NIR variability and the broad H$\alpha$ line observed in SBS~0335-052E may originate from a luminous blue variable outburst in a binary system. 
As part of their argument against an AGN interpretation, they showed that the observed [Fe {\sc ii}], [Fe {\sc iv}], and [Fe {\sc v}] emission lines cannot be reproduced by photoionization models with a pure AGN ionizing spectrum. 
However, their models assume AGN-only spectra. 
\citet{2024ApJ...966..170H} demonstrated that the ionizing spectrum of SBS~0335-052E is described by a combination of stellar and power-law components. 
While we do not rule out the possibility raised by \citet{2025arXiv250803912P}, an ionizing spectrum composed of both stellar and AGN components may explain the observed iron emission line ratio.

Hereafter, we assume that SBS~0335-052E hosts an AGN and discuss its BH properties.

\subsection{AGN Properites}

\subsubsection{A BH mass estimate based on the AGN bolometric luminosity}
\label{sec:BHmass}

\begin{figure}[t]
\plotone{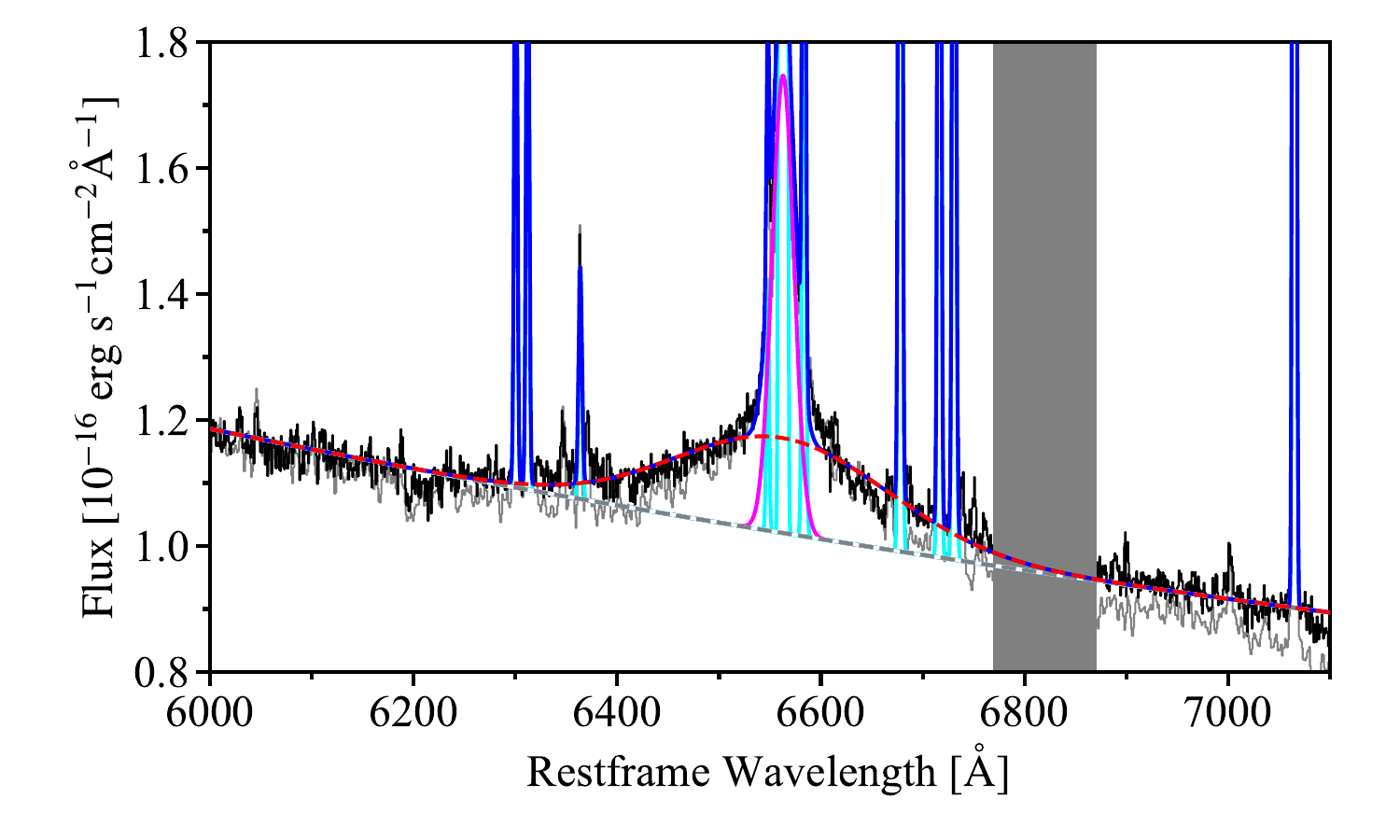}
\caption{Keck/LRIS optical spectrum of SBS~0335-052E obtained in 2021 (black line).
 The solid cyan, solid purple, and dashed red curves represent the best-fit emission lines of the narrow, medium, and broad components, respectively. The blue curve denotes the total fluxes of the best-fit emission lines. The grey shade indicates the spectrum contaminated by night sky emissions, which are masked out in our emission-line model fitting.
 The dashed grey line represents the best-fit continuum emission.
 The grey line shows the VLT/FORS optical spectrum taken in 2002 reported by \cite{2009AA...503...61I}, scaled so that its flux at 6500 \AA\ matches that of the Keck/LRIS spectrum for easy comparison.}
\label{fig:Ha}
\end{figure}

\begin{table}
\centering
\begin{tabular}{c c c}
\hline
\hline
Component & Flux [erg s$^{-1}$ cm$^{-2}$] & FWHM [km s$^{-1}$]\\
\hline
Narrow & $1.81 \ (\pm 0.002) \times 10^{-13}$ &$1.58\ (\pm 0.002)\times 10^{2}$\\
Medium & $2.40 \ (\pm 0.05)\times 10^{-15}$&$1.29 \ (\pm 0.06)\times 10^{3}$\\
Broad&$4.80 \ (\pm 0.1) \times 10^{-15}$&$1.24\ (\pm 0.08) \times 10^{4}$\\
\hline
\hline
\end{tabular}
\caption{Fluxes and line widths for the three components of the H$\alpha$ emission.}
\label{tab:Ha_flux}
\end{table}

We can infer a BH mass of SBS~0335-052E with the AGN bolometric luminosity estimated by the {\tt CIGALE} modeling in Section \ref{sec:AGN_scenario}.
As described in Section~\ref{sec:AGN_scenario}, the intrinsic AGN bolometric luminosity of SBS~0335-052E is estimated to be $L_{\rm bol} = 1.21\times 10^{43}$ erg s$^{-1}$ with the {\tt CIGALE} SED modeling.
The relation between the Eddington luminosity ($L_{\rm Edd}$) and the BH mass $M_{\rm BH}$ is given by $L_{\rm Edd} = 1.2\times 10^{38} (M_{\rm BH}/M_\odot)~{\rm erg \ s^{-1}}$, from which we get the BH mass of SBS~0335-052E as
\begin{equation}
M_{\rm BH} = 1.0\times 10^{5}~M_{\odot} \left(\frac{L_{\rm bol}}{L_{\rm Edd}}\right)^{-1} \left(\frac{L_{\rm bol}}{1.21 \times 10^{43}~{\rm erg~s}^{-1}}\right),
\label{eqn:eddington_luminosity}
\end{equation}
where $L_{\rm bol}/L_{\rm Edd}$ denotes the Eddington ratio.
If SBS~0335-052E is assumed to be accreting at the Eddington rate, its BH mass is estimated to be $M_{\rm BH} \simeq 10^{5}~M_{\odot}$, placing it in the IMBH mass range. 
More conservatively, given that dust torus NIR emission is typically observed in AGNs with Eddington ratios of $L_{\rm bol}/L_{\rm Edd} \sim 0.01$–$10$ \citep{2017ApJ...841...37G,1995ApJ...438L..37A,1988ApJ...332..646A}, the BH mass of SBS~0335-052E is expected to lie in the range $M_{\rm BH} \sim 10^{4}$–$10^{7}~M_{\odot}$.

\subsubsection{Virial BH mass inference from the (scattered) broad H$\alpha$ emission line}
\label{sec:BHmass_broadHa}

Interestingly, it has long been known that SBS~0335-052E (presumably its SSC~1+2 region) exhibits a broad H$\alpha$ component with full width at half maximum (FWHM) well in excess of $1,000~{\rm km~s}^{-1}$ in the optical spectrum \citep{2007ApJ...671.1297I,2009AA...503...61I}.
With the deep VLT/FORS spectrum of the SSC~1+2 region in SBS~0335-052E obtained in 2002, \citet{2009AA...503...61I} reported the presence of a broad component in the hydrogen lines H$\alpha$, H$\beta$, and H$\gamma$, but not in the strong forbidden lines.
They interpreted this as evidence for rapid motions of dense ionized gas with an electron number density of $N_{e} \gtrsim 10^{5-6}~\text{cm}^{-3}$, though without explicitly attributing it to AGN BLR emission.

In Figure~\ref{fig:Ha}, we compare the VLT/FORS spectrum from \citet[][their Fig.~5]{2009AA...503...61I} with the Keck/LRIS spectrum obtained in 2021 \citep{2024ApJ...966..170H} to examine the persistence of the broad H$\alpha$ line.
We clearly detect the broad H$\alpha$ line in the Keck/LRIS spectrum, with essentially the same shape as in the VLT/FORS spectrum, indicating that it has persisted for at least 19~years.
While SNe, particularly type IIn SNe, can produce broad H$\alpha$ lines with FWHM as large as $\sim 10{,}000~\mathrm{km~s^{-1}}$ \citep[e.g.,][]{2001MNRAS.326.1448C,2008ApJ...686..467S,2013A&A...555A..10T,2015MNRAS.449.1876S,2017ApJ...850...89G,2019ApJ...872..135K}, the flux and FWHM of such SN-originated broad H$\alpha$ lines are expected to vary on short timescales. 
As shown in Figure~\ref{fig:Ha}, the broad H$\alpha$ line in SBS~0335-052E persists for at least 19~years with no significant change, making a SN origin unlikely.

The broad H$\alpha$ line{, which} persists for $\gtrsim 19$~years can be naturally explained by emission from the AGN BLR. 
However, as already discussed above, SBS~0335-052E is likely to be a heavily-obscured AGN, which apparently contradicts the presence of the H$\alpha$ broad line from the AGN BLR. 
This discrepancy can be resolved if the broad component {is} the BLR emission that is scattered by electrons and/or dust in gaseous regions located along the polar direction of the AGN.
The presence of such scattering regions along the polar direction is a common component of AGNs, as evidenced by the detection of hidden broad lines in polarized light through spectropolarimetry of type 2 AGNs \citep[e.g.,][]{1985ApJ...297..621A,2003ApJ...583..632T,2016MNRAS.461.1387R,2022ApJ...937...65L}.
As will be shown below, the unusually large ratio of the SED-based AGN bolometric luminosity to the observed broad H$\alpha$ line luminosity, $L_{\rm bol}/L_{\rm H\alpha} \simeq 5760$, in contrast to the typical AGN value of $L_{\rm bol}/L_{\rm H\alpha} = 130$ \citep{2012MNRAS.423..600S}, provides strong evidence that the observed broad H$\alpha$ line in SBS~0335-052E is dominated by scattered rather than direct BLR emission.

In any case, if this broad H$\alpha$ line is identified as the AGN BLR emission, its luminosity and line width can, in principle, be used to estimate the virial BH mass.
To this end, we perform H$\alpha$ line profile fitting of the Keck/LRIS spectroscopic data, determining the line spread function (LSF) for the red channel of the instrument. 
We adopt a model for the LSF based on the O{\sc i} 5577 night-sky line, which we fit using two Gaussian components.
We perform line profile fitting to the H$\alpha$, [N {\sc ii}]$\lambda$6548, 6583, [O {\sc i}]$\lambda$6364 line and the other 6 weak emission lines ([O {\sc i}]$\lambda$6300, [S {\sc iii}]$\lambda$6312, He {\sc i}$\lambda$6678, [S {\sc ii}]$\lambda\lambda$6717,6731, and He {\sc i}$\lambda$7065 using the LSF with free parameters of line fluxes, widths, and central wavelengths, fitting the continuum emission using power-law with free parameters of a power-law index and an amplitude.
The faint emission lines (weaker than the [O {\sc i}]$\lambda$6364 line) and the strong night skylines that could contaminate the spectrum are masked during the line profile fitting.
{We find that there is a significant line broadening in the H$\alpha$ line of SBS 0335-052E. Although we include one additional line component for the H$\alpha$ line fitting, there remains a very broad component. We thus include one more component, and conduct the line profile fitting with three components dubbed narrow, medium, and broad components (Figure~\ref{fig:Ha}).}
The best-fit parameters are summarised in Table~\ref{tab:Ha_flux}.
We find that the FWHMs of the narrow, medium, and broad components are $158$, $1290$, and $12400~{\rm km~s}^{-1}$, respectively, in the best-fit profiles. 
The observed luminosity of the broad H$\alpha$ emission $L_{{\rm H}\alpha}$ is $L_{{\rm H}\alpha} = 2.10\ (\pm 0.04) \times10^{39}$ erg s$^{-1}$.

We estimate the virial BH mass assuming that the observed broad H$\alpha$ line originates from a scattering process.
Since the scattering by the gas surrounding the AGN is unlikely to be 100\% efficient, the intrinsic broad H$\alpha$ line should be brighter than the observed broad H$\alpha$ luminosity.
The scattering efficiencies $\epsilon$ of AGNs, defined as the ratio of the detected scattered flux density to the flux density we would detect if we had a direct view to the source, are typically $\epsilon \gtrsim\ $0.01 \citep[e.g.,][]{2005AJ....129.1212Z,2006AJ....132.1496Z}.
We adopt a lower limit of $\epsilon=0.01$ and derive the upper limit of the intrinsic H$\alpha$ luminosity as 
\begin{eqnarray}
L_{{\rm H}\alpha;{\rm corr}} &=& \epsilon^{-1} L_{{\rm H}\alpha} \nonumber\\
&=& 2.10~(\pm 0.04) \times10^{41}~\text{erg}~\text{s}^{-1}  \left(\frac{\epsilon}{0.01}\right)^{-1},
\label{eqn:scattered_luminosity}
\end{eqnarray}
where $L_{{\rm H}\alpha;{\rm corr}}$ denotes the intrinsic H$\alpha$ luminosity corrected for the scattering efficiency $\epsilon$.
As a reference, by adopting the empirical relation $L_{\rm bol} = 130 L_{{\rm H}\alpha} = 2.73\times10^{39}~{\rm erg~s}^{-1}$ \citep{2012MNRAS.423..600S} and comparing it with the SED-based value of $L_{\rm bol} = 1.21 \times 10^{43}~{\rm erg~s^{-1}}$ derived in Section~\ref{sec:AGN_scenario}, we obtain a representative estimate of the scattering efficiency of $\epsilon \simeq 0.023$ for SBS~0335-052E.

Moreover, because the scattering may broaden the H$\alpha$ line \citep[thermal broadening; e.g.,][]{2006ApJ...643..112L,2022ApJ...937...65L,2025arXiv250316595R}, the observed H$\alpha$ line width may be broader than the intrinsic broad line.
By denoting the broadening factor as $\zeta \geq 1$, the intrinsic H$\alpha$ line width corrected for the broadening is given by:
\begin{eqnarray}
\text{FWHM}_{{\rm H}\alpha;{\rm corr}} &=& \zeta^{-1} \text{FWHM}_{{\rm H}\alpha} \nonumber\\
&=& 12400~(\pm 800)~\text{km}~\text{s}^{-1}  \left(\frac{\zeta}{1}\right)^{-1},
\label{eqn:scattered_fwhm}
\end{eqnarray}
where $\mathrm{FWHM}_{{\rm H}\alpha}$ is the observed FWHM of the broad H$\alpha$ line (Table~\ref{tab:Ha_flux}), and $\mathrm{FWHM}_{{\rm H}\alpha;{\rm corr}}$ is the intrinsic FWHM corrected for the scattering-induced broadening.
Since a velocity width of $12400~{\rm km~s}^{-1}$ is exceptionally large for a typical AGN BLR, it is likely that $\zeta \gg 1$.
If we somewhat boldly assume that the medium-width H$\alpha$ component (Table~\ref{tab:Ha_flux}) represents the intrinsic FWHM of the AGN BLR in SBS~0335-052E, we obtain $\zeta = 12400 / 1290 \simeq 9.61$.

By using the empirical single-epoch virial BH mass estimator established for Type 1 AGNs \citep{2005ApJ...630..122G}, we get from Equations~\ref{eqn:scattered_luminosity} and \ref{eqn:scattered_fwhm}:
\begin{eqnarray}
M_{\text{BH}} & = & 2.0^{+0.4}_{-0.3} \times 10^{6}~M_{\odot} \nonumber\\
& \times &\left( \frac{L_{\text{H}\alpha;\text{corr}}}{10^{42}~\text{erg}~\text{s}^{-1}} \right)^{0.55 \pm 0.02} \left( \frac{\text{FWHM}_{\text{H}\alpha;\text{corr}}}{1000~\text{km}~\text{s}^{-1}} \right)^{2.06\pm 0.06} \nonumber\\
&\simeq& 1.5~(\pm 0.4) \times 10^{8}~M_{\odot} \nonumber\\
&\times& \left(\frac{\epsilon}{0.01}\right)^{-0.55\pm0.02} \left(\frac{\zeta}{1}\right)^{-2.06\pm0.06}.
\label{eqn:mass_estimator}
\end{eqnarray}
Thus, we give a stringent upper limit to the BH mass of SBS~0335-052E as $M_{\rm BH} < 1.5~(\pm 0.4)\times 10^{8} M_\odot$.
If we adopt the above-mentioned bold estimates of $\epsilon = 0.023$ and $\zeta = 9.61$, Equation~\ref{eqn:mass_estimator} yields $M_{\rm BH} \simeq 9.0~(\pm 2.4) \times 10^{5}~M_\odot$.

Although here we assume that the broad H$\alpha$ line in SBS~0335-052E originates from a AGN BLR, the H$\alpha$ observations alone cannot completely exclude other possibilities, such as production of a broad H$\alpha$ component by multiple SNe \citep[e.g.,][]{2007ApJ...671.1297I}, extremely long-lasting Type~IIn supernovae with complex CSM structures \citep[e.g.,][]{2014ApJ...797..118F,2016MNRAS.458.2063K,2017A&A...605A...6N,2019ApJ...872..135K,2020MNRAS.491.1384M}, or other stochastically variable events such as luminous blue variables \citep[e.g.,][]{2014MNRAS.445..515K,2022MNRAS.515..110K,2025arXiv250803912P}.
Note that a persistent broad hydrogen emission lines could also be produced by the hydrogen Raman scattering \citep[e.g.,][]{2016ApJ...824L..13D,2024MNRAS.529.2131K,2024arXiv240704777K}.

\subsubsection{$M_{\rm BH}-M_{*}$ Relation}
\label{sec:mstar_mbh}

\begin{figure*}[t]
\plotone{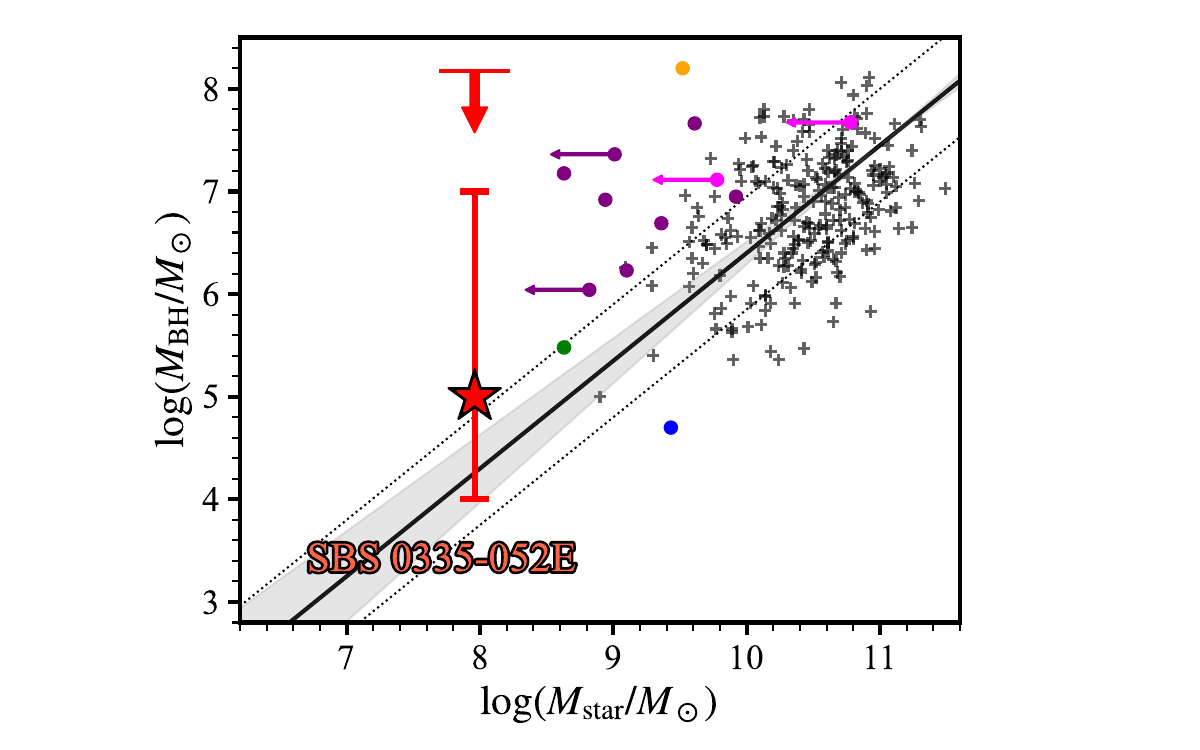}
\caption{Stellar mass to BH mass relation.
 The red star mark represents the BH mass $M_{\rm BH}$ (Equation~\ref{eqn:eddington_luminosity}) and the total stellar mass $M_{*}$ of SBS~0335-052E, inferred from our best-fit SED.
 The downward arrow indicates the upper limit of the BH mass derived from the spectral fitting for the emission line (Equation~\ref{eqn:mass_estimator}).
 The black crosses are the BH masses and total stellar masses of local broad-line dwarf galaxies taken from \citep{2015ApJ...813...82R}. 
 The blue and green circles represent the least-massive local AGNs RGG~118 \citep{2015ApJ...809L..14B} and POX~52 \citep{2008ApJ...686..892T}, respectively.
 The magenta, orange, and purple circles denote {\it JWST}-detected broad-line galaxies at $z \gtrsim 4$ identified by \cite{2023arXiv230200012K}, \cite{2023A&A...677A.145U}, and \cite{2023arXiv230311946H}, respectively, where we omit two objects of \cite{2023arXiv230200012K} from the data points of \cite{2023arXiv230311946H}.
 The black solid line and the grey shade show the best-fit line and $1\sigma$ error of the local relation in the range of $M_{\rm BH}=10^5-10^{8.5} M_\odot$ provided by \cite{2015ApJ...813...82R}. 
 The dotted lines present the sum of intrinsic scatter and measurement uncertainty in the local relation.}
 \label{fig:MsMb}
\end{figure*}

Figure~\ref{fig:MsMb} shows the BH mass vs. host galaxy stellar mass plot, in which SBS~0335-052E is compared with nearby type 1 AGNs \citep{2015ApJ...813...82R}, including the least massive known AGNs POX~52 \citep{2008ApJ...686..892T} and RGG~118 \citep{2015ApJ...809L..14B}, as well as with the {\it JWST}-discovered high-$z$ ($z \gtrsim 4$) broad-line AGNs \citep[][see Section~\ref{sec:jwst} for more details]{2023arXiv230200012K,2023A&A...677A.145U,2023arXiv230311946H}.
For SBS~0335-052E, we adopt the total stellar mass derived from the best-fit SED described in Section~\ref{sec:AGN_scenario}, a BH mass range of $M_{\rm BH}=10^{4}$–$10^{7}~M_{\odot}$ with a central value of $M_{\rm BH}=1.0 \times 10^{5}~M_{\odot}$ (Equation~\ref{eqn:eddington_luminosity}), and an upper limit of $M_{\rm BH}=1.5 \times 10^{8}~M_{\odot}$ (Equation~\ref{eqn:mass_estimator}).

{
{We {can} see in Figure~\ref{fig:MsMb} that SBS 0335-052E occupies the lowest stellar-mass regime among the known AGN {samples}.}
The host stellar mass of SBS~0335-052E is significantly less massive than that of the nearby AGNs and the {\it JWST}-discovered broad-line AGNs with host stellar mass estimates.
There are no known broad-line AGNs whose properties are equivalent to those of SBS~0335-052E known to date.

{{Considering} its $M_{\rm BH}$ together with the {host} stellar mass, SBS~0335-052E is consistent with, or lies above, the extrapolated local $M_{\rm BH}-M_{*}$ relation.}
{
{However}, the current uncertainty in $M_{\rm BH}$ of SBS~0335-052E is too large
{
{to determine whether its central BH is {truly} overmassive or consistent with the relation}, 
{
{preventing us from placing} meaningful constraints on seed BH formation scenarios \citep[e.g., see][and references therein]{2020ARA&A..58..257G}. 
Future observations -- such as high spatial resolution integral field spectroscopy capable of resolving the galaxy’s BH sphere of influence to analyze stellar or gas dynamics \citep[e.g.,][]{2025arXiv250821748J} -- will be crucial for tightly constraining $M_{\rm BH}$ in SBS~0335-052E and for investigating possible seed BH formation mechanisms.

\subsection{Relation to {\it JWST}-detected High Redshift Broad-line AGNs}
\label{sec:jwst}

If SBS~0335-052E is interpreted as hosting an AGN, it can be regarded as a local analog of low-mass galaxies at high redshifts that harbor AGNs. 
Although Figure~\ref{fig:MsMb} shows that the stellar masses of the {\it JWST}-detected high-$z$ broad-line AGNs are not as low as that of SBS~0335-052E (as discussed in Section~\ref{sec:mstar_mbh}), and their metallicities ($Z=0.2$–$0.4~Z_{\odot}$) are mostly higher than that of SBS~0335-052E (e.g., \citealt{2023arXiv230311946H, 2023arXiv230200012K}, but see also \citealt{2025arXiv250522567M}), it remains instructive to compare SBS~0335-052E with these high-$z$ AGNs, highlighting both similarities and differences.

{\it JWST} observations have recently revealed a large population of high-redshift AGNs at $z \gtrsim 4$ exhibiting broad H$\alpha$ (and sometimes H$\beta$) emission lines (e.g., \citealt{2023ApJ...954L...4K,2023A&A...677A.145U, 2023arXiv230311946H,2023ApJ...957L...7K,2024A&A...691A.145M,2024MNRAS.531..355U,2024ApJ...963..129M, 2024Natur.628...57F, 2024ApJ...964...39G, 2024Natur.636..594J,2025ApJ...986..126K}).
While outflows can produce broad H$\alpha$, in such cases forbidden lines (e.g., [O {\sc iii}]$\lambda\lambda$4959,5007) with comparable widths and strengths are typically observed \citep{2019ApJ...873..102F,2022ApJ...929..134X}.
However, the {\it JWST}-detected high-$z$ broad-line AGNs do not show forbidden lines as broad as H$\alpha$, suggesting that the broad H$\alpha$ is unlikely to originate from outflows and is more naturally explained by emission from the broad-line region around the central AGN, whose SMBH masses are overmassive \citep[$M_{\rm BH}/M_{*} \gtrsim 0.01$ as shown in Figure~\ref{fig:MsMb};][and references therein]{2025arXiv250821748J,2025arXiv250403551J}.
Interestingly, a significant fraction ($\gtrsim 20$\%) of the {\it JWST}-detected broad-line AGNs are classified as so-called Little Red Dots (LRDs), which are characterized by a V-shaped SED in the rest-frame UV-optical range, likely caused by a deep dip at the hydrogen Balmer edge (3647\AA) produced by absorption from dense neutral gas surrounding the central AGN \citep[e.g.,][]{2024ApJ...963..129M,2024ApJ...964...39G,2024ApJ...968....4P,2025ApJ...986..126K}.
Although these {\it JWST}-detected high-$z$ broad-line AGNs exhibit broad Balmer emission lines (i.e., indicative of unobscured AGNs), they show rest-frame hard X-ray luminosities far weaker than expected by a factor of $\gtrsim 10$ if they were truly unobscured AGNs \citep[e.g.,][]{2024ApJ...969L..18A,2024arXiv240704777K,2025MNRAS.538.1921M}.

The observational property of exhibiting broad H$\alpha$ emission while lacking correspondingly strong X-ray emission, as seen in the {\it JWST}-detected broad-line AGNs, is similar to the case of SBS~0335-052E (see Section~\ref{sec:appendix})\footnote{After the submission of our paper to arXiv, \cite{2024MNRAS.535..853J} point out that the a broad-line LRD GN-280974 at $z = 2.26$ {particularly resembles} SBS~0335-052E in terms of the X-ray and radio weakness, claiming that SBS~0335-052E is the closest and clearest example of low-$z$ analog of the {\it JWST}-detected broad-line AGNs.}.
The X-ray weakness of the {\it JWST}-detected broad-line AGNs is generally interpreted as the result of Compton-thick (dust-free) gas absorption due to gases with $N_{H}\gtrsim 10^{24}~\text{cm}^{-2}$ surrounding the AGNs (e.g., \citealt{2024ApJ...969L..18A,2024arXiv241203653I,2025MNRAS.538.1921M}, but see also \citealt{2024arXiv240704777K,2024ApJ...977L..13B} for alternative explanations).
This interpretation is consistent with the approach we adopt to explain the X-ray weakness of SBS~0335-052E (Section~\ref{sec:appendix}).

However, we should note that SBS~0335-052E and the {\it JWST}-detected broad-line AGNs also show notable differences in their observed properties.
In particular, the latter exhibit no detectable IR dust–torus emission \citep[e.g.,][]{2024ApJ...968...34W,2024ApJ...968....4P,2025ApJ...991...37A}, in sharp contrast to SBS~0335-052E, where very strong and variable IR hot dust emission is clearly observed.
Moreover, the UV–optical SED of SBS~0335-052E lacks the characteristic V-shaped profile seen in the LRD broad-line population (although it does not differ significantly from the UV–optical SEDs of the non-LRD broad-line population).
Regarding the broad H$\alpha$ line width, the FWHM of $\gtrsim 10000~\text{km~s}^{-1}$ observed in SBS~0335-052E is much greater than that of the typical {\it JWST}-detected broad-line AGNs \citep[$1000-5000~\text{km~s}^{-1}$;][]{2025MNRAS.538.1921M}, which may suggest that scattering-induced broadening is much more significant in the former than in the latter \citep{2025arXiv250316595R}.

The possible analogy between SBS~0335-052E and the {\it JWST}-detected broad H$\alpha$ emitters is certainly intriguing and warrants further investigation; however, any analogy based solely on similarities in some of the observational properties (e.g., the X-ray weakness despite the presence of broad emission lines) should be treated with caution.

\subsection{Possible Dual AGN in SBS~0335-052E}

In this paper, we have assumed that the {\it WISE}-detected variability in SBS~0335-052E originates from a single source. 
However, it is worth noting that the observed {\it WISE} variability could, in principle, be driven by multiple AGNs.

The high-resolution NIR continuum image of SBS~0335-052E taken with the {\it HST} resolves the {\it WISE} source into multiple NIR components, including two major hot-dust sources, SSC~1 and SSC~2, which are separated by a projected distance of $\sim 100$~pc on the sky \citep{2008AJ....136.1415R}.
SSC~1 and SSC~2 have similar stellar masses ($M_* \sim 1\times 10^6 M_\odot$) and ages ($\lesssim$ 3 Myr), and both of them exhibit the IR excess emissioin \citep[SSC~1 is about 1.4 times brighter than SSC 2 at $\sim 2~\mu$m;][]{2008AJ....136.1415R}.
We note that due to the low spatial resolution of {\it WISE}, it is unclear which of the two SSCs actually contributes to the WISE-detected variability (or whether both do).

Adopting our interpretation of the IR excess as AGN-heated hot dust, both SSC~1 and SSC~2 can be interpreted as AGNs, each potentially harboring a massive BH.
Because SSC~1 and SSC~2 are separated by only $\sim$100 pc, SBS~0335-052E could represent a candidate dual AGN system in the process of merging. 
If confirmed, this system is one of the closest separation dual AGN \citep[e.g., see][and references therein]{2019ApJ...879L..21G,2023ApJ...942L..24K,2024ApJ...972..185T,2025ApJ...989..112G}.

Obscured dual AGNs in dwarf galaxies are important for understanding merger-triggered accretion, the hierarchical growth of structures, and the generation of gravitational waves \citep[e.g.,][]{2015ApJ...806..219C,2017ApJ...848..126S,2019ApJ...879L..21G,2023ApJ...944..160M}.
A detailed confirmation of such a system in SBS~0335-052E is therefore crucial and warrants further investigation, which is beyond the scope of this paper but will be addressed elsewhere.

\section{Summary}
\label{sec:conclusions}

In this work, we report the discovery of the NIR variability in the star-forming blue compact dwarf galaxy SBS~0335-052E and interpret the variable NIR emission as originating from the AGN dust torus associated with a Compton-thick, heavily obscured AGN.
Our main findings are summarized as follows:

\begin{enumerate}
\item We analyzed the {\it WISE} W1 and W2 band light curves of SBS~0335-052E spanning 14~years, carefully correcting for the systematic offsets (Section~\ref{sec:result}). We find that both the $W1$ and $W2$ fluxes exhibit multiple rising and declining events on timescales of several years, which cannot be explained by one-off transient phenomena or by emission from an extended ($\gtrsim 1$~pc) hot dust distribution. All of the observed NIR variability properties are consistent with emission from an AGN hot dust torus.
\item We performed UV-IR SED fitting with {\tt CIGALE} using archival broad-band photometry of SBS~0335-052E (Section~\ref{sec:AGN_scenario}). We find that while the optical spectrum is dominated by stellar light, the NIR emission is well reproduced by an edge-on, heavily obscured AGN dust torus SED model. The W1-W2 FVG analysis independently confirms the Compton-thick nature of the obscuration of the AGN in SBS~0335-052E ($N_{\rm H} \gtrsim 10^{24}~{\rm cm}^{-2}$; Section~\ref{sec:appendix}). The observed low X-ray luminosity in SBS~0335-052E is consistent with the Compton-thick AGN interpretation.
\item We find that the H$\alpha$ emission line profile of SBS~0335-052E consists of narrow, intermediate, and broad components (Section~\ref{sec:BHmass_broadHa}). The broad component has a FWHM of $\sim$12,000~km~s$^{-1}$. A comparison between the VLT/FORS and Keck/LRIS spectra, obtained $\sim$19~years apart, shows that the H$\alpha$ profile has remained stable over this period, with no significant changes in either flux or FWHM. If the broad component originates from the AGN BLR, its detection despite the inferred Compton-thick obscuration appears contradictory. We therefore interpret the broad H$\alpha$ component as AGN BLR emission scattered by electrons (or dust grains) located along the polar direction of the AGN. The unusually large ratio of the SED-based AGN bolometric luminosity to the observed broad H$\alpha$ line luminosity, $L_{\rm bol}/L_{\rm H\alpha} \simeq 5760$, compared to the typical AGN value of $L_{\rm bol}/L_{\rm H\alpha} = 130$, further supports the scattered BLR interpretation.
\item From the SED analysis, the AGN bolometric luminosity of SBS~0335-052E is estimated to be $L_{\rm bol} = 1.21\times 10^{43}~{\rm erg~s}^{-1}$, with which we obtain $M_{\rm BH} \simeq 10^{5}~M_{\odot}$ assuming the Eddington accretion rate (Equation~\ref{eqn:eddington_luminosity}). Also, by taking into account the scattering efficiency and line broadening due to the scattering, we obtain an upper limit on $M_{\rm BH}$ from the observed broad H$\alpha$ line luminosity and FWHM as $M_{\rm BH} < 1.5\times 10^{8}~M_{\odot}$ (Equation~\ref{eqn:mass_estimator}). Although the estimate of $M_{\rm BH}$ carries large uncertainties, the AGN in SBS~0335-052E (if confirmed) would be the one hosted by the lowest-mass galaxy known, and may be powered by an IMBH (Section~\ref{sec:mstar_mbh}).
\end{enumerate}


\begin{acknowledgments}

We thank K. Shimasaku, T. Tanaka, S. Huang, and R. Kawabe for useful discussions.

This research is based in part on data gathered with the 10-meter Keck Telescope located at W. M. Keck Observatory. 
We thank the observatory personnel for their help with the observations.
This work was supported by the joint research program of the Institute of Cosmic Ray Research (ICRR), the University of Tokyo. 
    This work was supported by the World Premier International Research Center Initiative (WPI Initiative), MEXT, Japan, as well as KAKENHI Grant-in-Aid for Scientific Research (24KJ1159, 19H00697, 20H00180, 21H04467, 21K03622, 24K17097, and {25K07370}) through the Japan Society for the Promotion of Science (JSPS). This work was partially supported by Overseas Travel Fund for Students (2024) of Astronomical Science
Program, The Graduate University for Advanced Studies, SOKENDAI.
This publication makes use of data products from the Wide-field Infrared Survey Explorer, which is a joint project of the University of California, Los Angeles, and the Jet Propulsion Laboratory/California Institute of Technology, funded by the National Aeronautics and Space Administration.

\end{acknowledgments}

%

\vspace{5mm}
\facilities{WISE, NEOWISE}


\software{
        astropy \citep{2013A&A...558A..33A,2018AJ....156..123A,2022ApJ...935..167A},  
        linmix\_err \citep{2007ApJ...665.1489K},
        CIGALE v.2022.1 \citep{2020MNRAS.491..740Y},
        IRAF \citep{1993ASPC...52..173T,1986SPIE..627..733T},
        WISE/NEOWISE Coadder ICORE \citep{2009ASPC..411...67M,2013ascl.soft02010M},
        XSPEC~v.12.14.1 \citep{1996ASPC..101...17A}.
          }




\bibliography{hatano25}

\end{document}